\documentclass[lettersize,journal]{IEEEtran}
\IEEEoverridecommandlockouts
\usepackage{textcomp}
\usepackage{xcolor}
\usepackage{tcolorbox}
\usepackage{cite}
\usepackage{hyperref}
\usepackage{balance}
\usepackage{amsmath,amssymb,amsfonts}
\usepackage{algorithmic}
 \usepackage{url}
\usepackage{graphicx}
\usepackage{float}
\usepackage{textcomp}
\usepackage{xcolor}
\usepackage{multirow}
\usepackage{rotating}
\usepackage{booktabs}
\def\BibTeX{{\rm B\kern-.05em{\sc i\kern-.025em b}\kern-.08em
    T\kern-.1667em\lower.7ex\hbox{E}\kern-.125emX}}

\begin{document}
\title{Comparing Self-Disclosure Themes and Semantics to a Human, a Robot, and a Disembodied Agent}

\author{Sophie Chiang$^{1,\dagger}$, Guy Laban$^{1,\dagger,*}$, Emily S. Cross$^{2}$, Hatice Gunes$^{1}$\\
 $^1$Department of Computer Science and Technology, University of Cambridge, Cambridge, United Kingdom\\
 $^2$Professorship for Social Brain Sciences, ETH Zürich, Switzerland
\thanks{$\dagger$ Equal Contribution}\thanks{* Corresponding author {\tt\small guy.laban@cl.cam.ac.uk}}
\thanks{G. Laban, and H. Gunes have been supported by the EPSRC project ARoEQ under grant ref. EP/R030782/1. G. Laban and E. S. Cross were funded by the European Union’s Horizon 2020 Research and Innovation Programme under the Marie Skłodowska-Curie to ENTWINE, the European Training Network on Informal Care (Grant agreement no. 814072). E. S. Cross is funded by the European Research Council (ERC) under the European Union’s Horizon 2020 Research and Innovation Programme (Grant agreement no. 677270 to EC), and the Leverhulme Trust (PLP-2018-152 to EC).}
\thanks{\textbf{Open Access:} For the purpose of open access, the authors have applied a Creative Commons Attribution (CC BY) licence to any Author Accepted Manuscript version arising. 
}}

\maketitle

\begin{abstract}

As social robots and other artificial agents become more conversationally capable, it is important to understand whether the content and meaning of self-disclosure towards these agents changes depending on the agent’s embodiment. In this study, we analysed conversational data from three controlled experiments in which participants self-disclosed to a human, a humanoid social robot, and a disembodied conversational agent. Using sentence embeddings and clustering, we identified themes in participants’ disclosures, which were then labelled and explained by a large language model. We subsequently assessed whether these themes and the underlying semantic structure of the disclosures varied by agent embodiment. Our findings reveal strong consistency: thematic distributions did not significantly differ across embodiments, and semantic similarity analyses showed that disclosures were expressed in highly comparable ways. These results suggest that while embodiment may influence human behaviour in human–robot and human–agent interactions, people tend to maintain a consistent thematic focus and semantic structure in their disclosures, whether speaking to humans or artificial interlocutors.

\end{abstract}

Advancements in conversational technologies will enable social robots and other conversational artificial agents to take part in an increasing range of verbal social interactions, 
from everyday conversations to more structured exchanges \cite{Henschel2021,Kim2024UnderstandingInteraction}. One critical aspect of these interactions is self-disclosure—the process by which individuals reveal personal thoughts, feelings, and experiences \cite{RefWorks:357}. In both human--human and human--robot interactions, self-disclosure shapes relational dynamics and behaviour, as well as the way individuals interpret and make sense of their own experiences \cite{Laban2024SharingFeel}. Nonetheless, a gap remains in our understanding of how self-disclosure differ thematically and semantically towards social robots, compared to artificial agents of varying levels of embodiment. Previous studies on verbal communication and self-disclosure in human–robot interaction (HRI) have primarily focused on the quantity of disclosure, expressive behaviour, and the emotional drivers and outcomes \cite{Laban2021TellSpeech,Laban2024BuildingTime,laban_ced_2023,2023OpeningBehavior}. 
However, little attention has been given to exploring whether the thematic and semantic content of self-disclosure remains consistent when individuals engage with agents of varying embodiments, ranging from human interlocutors to voice assistants and physical robots.

The themes individuals disclose provides an insight into their lives and experiences \cite{RefWorks:455}, providing a window for understanding what matters to them 
during interpersonal communication \cite{Berg1987ThemesSelf-Disclosure}. The topics people choose to disclose often reflect social norms, emotional needs, and expectations about the listener’s role 
\cite{Collins1994Self-disclosureReview}. 
Understanding whether thematic structures remain stable or shift based on an agent’s embodiment can reveal whether people treat social communication to social robots and other artificial agents based on their physical form, potentially shaping the themes of information users share. Investigating these thematic patterns can provide insights into whether self-disclosure in HRI is consistent with the same underlying principles as human interaction, or if it is uniquely shaped by agent embodiment. If thematic consistency is observed, it would suggest that self-disclosure content is primarily driven by individual tendencies and broader social factors rather than the embodiment of the agent itself. Conversely, if the themes of disclosure differ across agent embodiments, it could indicate that embodiment influences not only the interaction experience and verbal performance \cite{Laban2021TellSpeech,spitale_hri_,2021PerceptionsIntelligence,2022DontAgents,Laban2020TheSystems} but also what individuals decide sharing. Accordingly, we are asking \textbf{(RQ1)}: \textit{To what extent is self-disclosure to artificial agents of varying levels of embodiment thematically consistent?}

Beyond thematic content, the way self-disclosure is formulated, its semantic structure, shapes how meaning is conveyed and could be understood in communication \cite{Cunningham2006MeaningReference}. 
The degree to which individuals alter their wording and meaning across different conversational partners can provide insight into how they manage social communication and tailor it to diverse contexts and interlocutors \cite{Gregory1967AspectsDifferentiation}.
In HRI, assessing semantics in self-disclosure can help determine whether robotic embodiment impacts not just \textit{what} is shared but \textit{how} its meaning is expressed. If semantic shifts occur, this may indicate that different embodiments elicit different cognitive and social framing of disclosure. Accordingly, we are asking \textbf{(RQ2)}: \textit{To what extent is self-disclosure to artificial agents of varying levels of embodiment semantically consistent?}

To answer our research questions, we conducted a secondary analysis of self-disclosure themes across diverse interaction contexts, including human-to-human, human-to-robot, and human-to-agent communications. We employ clustering techniques to group similar disclosures towards these agents and use a Large Language Model (LLM) to label and explain these clusters, thereby uncovering underlying thematic patterns. We further analyse the distribution of responses, represented by ratios across different agent embodiments, to quantify the prevalence of these themes and to assess whether individuals’ disclosure topics are influenced by the agent's embodiment. Finally, by applying semantic similarity metrics, we evaluate the consistency of conversational content across these interactions. Overall, our study aims to validate the persistence of shared communicative themes and meaning despite differences in agent embodiment.

\section{Related Work}

Communication Accommodation Theory (CAT) \cite{Giles1991ContextIntraction} provides a broad framework to understand how people may adjust their language to their conversation partners. It posits that individuals modify their communicative behaviour in interaction to align with or diverge from their conversation partner’s style for social reasons. Such adjustments can be linguistic, paralinguistic, or nonverbal – for example, people might change their word choice, speaking rate, accent, prosody, or even posture to become more similar to (i.e., \textit{convergence}) or deliberately different from (i.e., \textit{divergence}) their interlocutor \cite{shepard2001communication}. CAT argues that these shifts are often made to gain social approval, increase understanding, or manage social distance between speakers \cite{Giles1991ContextIntraction}. Psycholinguistic research has shown that conversation partners often unconsciously synchronize aspects of their language use. 
In dyadic interactions, people tend to match each other’s use of function words, sentence structures, or emotional language \cite{Gonzales2010LanguageGroups}. Speakers adjust their register, the level of formality or style, based on the situation and interlocutors \cite{Gregory1967AspectsDifferentiation}. People tend to choose different conversation topics or use different vocabulary, tone, and grammar when, for example, narrating a story versus giving instructions, or when speaking in a formal setting versus a casual one \cite{Pescuma2023SituatingMethods}.

When interacting with social robots and other 
artificial agents, humans might bring similar adaptive tendencies to bear, often treating the machine as a social interlocutor to some degree \cite{Laban2024SharingFeel}. Research in HRI communication shows many parallels to human-human accommodation. For example, one previous study shows that when self-disclosing to a robot (NAO, SoftBank Robotics), people were accommodating and spoke with a higher vocal pitch, adapting to its child-like embodiment. On the other hand, when speaking to a disembodied conversational agent (Google Nest Mini), their vocal harmonicity was higher, indicating a clearer voice, free from cracks and breaks \cite{Laban2021TellSpeech}. Other studies support that, showing that speech directed towards a voice assistant (Siri, Apple and Alexa, Amazon) is often louder and slower, suggesting that the disembodiment of such agents introduce speakers to intelligibility barriers \cite{Cohn2022Acoustic-phoneticSpeech}.

On the sociolinguistic level, a previous study reports that the sentiment of the content is influenced by the disclosure topic rather than the agent's embodiment \cite{Laban2021TellSpeech}. Lukin et al. \cite{Lukin2018ConsequencesDialogue} observed spontaneous variation in instruction-giving to a robot, identifying differences in verbosity and multi-intent structuring. These stylistic choices were shaped by individual user traits, trust in the robot, and accumulated interaction experience. This is in line with previous results showing that people who report for experiencing negative emotions (e.g., loneliness, stress, and low mood) tend to disclose more towards robots \cite{2023OpeningBehavior}. Similarly, Irfan and Skantze \cite{IrfanBetweenInteraction} showed that people tend to share personal stories with a robot 
in 
emotionally resonant contexts like moral dilemmas. 
A study by Asano et al. \cite{Asano2022ComparisonInteractions} reported that participants teaching a robot tended to align their word usage with the robot’s vocabulary, finding greater lexical alignment in one-on-one HRIs than in triads where an additional human was also present. 

When these adaptive tendencies extend into HRI, users often simplify, clarify, or modulate their language for intelligibility or to compensate for perceived limitations of the robot interlocutor \cite{Laban2021TellSpeech,Cohn2022Acoustic-phoneticSpeech}. Accordingly, in single session interactions people tend to share less with robots or agents than with fellow humans while often being aware of modulating their disclosures \cite{Laban2021TellSpeech}. However, when interactions extend beyond a single session, people who interact repeatedly with a robot show increasing self-disclosure and even report the robot to seem more socially capable over time \cite{Laban2024BuildingTime,laban_ced_2023}. Users often adjust their communication depending on their perception of the robot's communicative abilities. For instance, Skantze and Irfan \cite{10.5555/3721488.3721593} demonstrate that users interacting with a robot employing sophisticated turn-taking models (vs. a basic silence-threshold baseline) experienced fewer interruptions, shorter response times, and engaged in more natural conversational dynamics. 
However, no work to date has investigated the thematic framing and semantic structure of self-disclosure across different agent embodiments, furthering our understanding of the unique cognitive and social orientations people adopt when engaging with social robots. This line of research moves beyond measuring disclosure volume or affective behaviour, offering insights into how meaning-making unfolds in human–agent communication at the level of the utterance.

\section{Method}

This work includes a secondary analysis of a dataset collected from three distinct laboratory experiments (see Section \ref{sec:data}) carried out in previous research (see \cite{Laban2021TellSpeech}), aiming to evaluate the consistency of themes and semantics in self-disclosure to social (human) and artificial agents (a social robot and a disembodied agent) with varying levels of embodiment. Responses from all subjects' disclosure data \cite{Laban2021TellSpeech}, including all agent types (see Section \ref{sec:data}), were clustered to identify themes of disclosure (see Section \ref{subsec:clustering_pipeling}). To further understand the content within each cluster (theme of disclosure), we used LLM (GPT 4o-mini) for generating descriptions and labels for the clusters based on the content in participants disclosures (see Section \ref{sec:llm}), which were validated accordingly (see Section \ref{method:validation}). Then, we conducted statistical analysis to evaluate the thematic disparity across the three embodiments (human, humanoid social robot, and a disembodied artificial agent), as well as semantic similarity using various embedding models to evaluate the shared meaning in participants' disclosures to the three agents (see Section \ref{method:analysis}).

\subsection{Dataset}
\label{sec:data}

\begin{figure}[h!]
    \centering
    \includegraphics[width=1\linewidth]{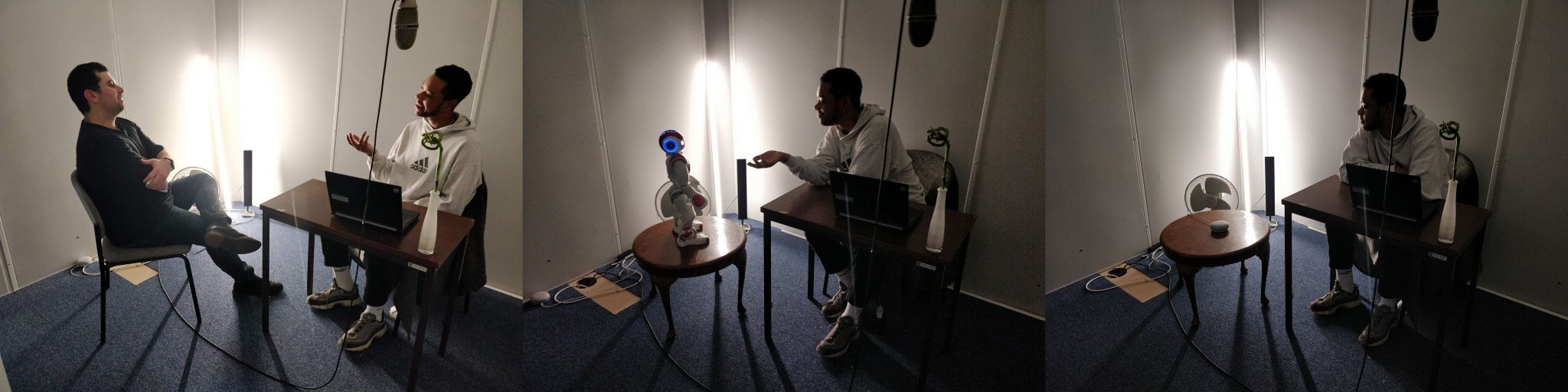}
    \caption{Illustration of the experimental design from \cite{Laban2021TellSpeech}. From left to right: human talking to a human agent, 
    to the social robot (NAO), and 
    to the disembodied agent (Google Nest Mini).}
    \label{fig:stimuli}
\end{figure}

The data collection of the data used in this study is reported in \cite{Laban2021TellSpeech}. 
The data were collected through three separate within-subjects laboratory experiments (i.e, user studies) where participants interacted with a humanoid social robot (NAO), a human, and a disembodied conversational agent (Google Nest Mini), each asking questions designed to elicit self-disclosure (see Figure \ref{fig:stimuli}). In Experiment 1, 26 participants each answered one question per agent, resulting in 78 recorded disclosures covering topics relevant to student life: academic assessment, student finances, and university–life balance. In Experiment 2, 
27 
participants, each responding to two questions per agent, yielding 162 data points across topics grouped into work and finances, 
social life and leisure, 
and intimate and family relationships. 
In Experiment 3, 61 participants participated in the same 
design, producing 366 disclosures across revised topics of work–life balance, 
relationships and social life, 
and physical and mental health. 
A Wizard-of-Oz setup was used to ensure consistent, pre-scripted interactions across the artificial agents, and the human interlocutor followed a rigorous protocol to ensure a systematic and valid comparison. All interactions were randomized and recorded in a soundproof lab \cite{Laban2021TellSpeech}. Participant disclosures towards the three agents, human-to-human (H2H), human-to-disembodied agent (H2A), and human-to-robot (H2R), were combined and labelled accordingly.

\subsection{Preprocessing and Clustering}
\label{subsec:clustering_pipeling}
Each data unit (i.e., an entire disclosure communicated by a participant to each of the three agents) was first filtered for stop words, 
after splitting the text by whitespace. Duplicate responses were removed prior to analysis. The resulting texts were converted into 384-dimensional sentence embeddings using the \textit{``all-MiniLM-L6-v2 model"} \cite{10.5555/3495724.3496209} from the SentenceTransformers library \cite{Reimers2019Sentence-BERT:BERT-Networks}. These embeddings were clustered using the K-means algorithm \cite{likas2003global}, with a fixed random seed 
to ensure replicability. The optimal number of clusters was selected via the elbow method \cite{liu2020determine}, based on inertia values. Each disclosure was subsequently assigned to one of the resulting clusters based on its semantic embedding.


\subsection{Cluster's Explanation}
\label{sec:llm}

Following the clustering of disclosures, 
the top $n$ disclosures closest to the centroid of a given cluster were selected by calculating the distance between response embeddings and their respective centroid. These were then concatenated into a prompt, reducing variance in the generated descriptions while also utilising the response embeddings. Accordingly, GPT 4o-mini, a state-of-the-art language model chosen for this task due to its fast and powerful performance, was provided with the following prompt to explain the clusters by assigning each a label and a detailed description:
\begin{tcolorbox}
\textit{\small “The following are responses to questions from a specific cluster. Analyse these responses and provide: 1) A concise label summarizing the main theme or central topic of this cluster; 2) A detailed paragraph describing key themes, patterns, or insights. Highlight any notable 
trends specific to this cluster.
Do not include any introductory statements or additional commentary.
Only provide the label and description, without introductory statements or commentary.”}
\end{tcolorbox}

\subsection{Validation}
\label{method:validation}

To validate the LLM-generated descriptions of the clusters as a sanity check
cosine similarity was calculated using ``\textit{all-MiniLM-L6-v2 model}" \cite{10.5555/3495724.3496209} 
for all pairwise combinations between the embeddings of each description and its corresponding cluster centroid with the following equation: 

\begin{equation*}\small
\max \left( \frac{\mathbf{D}_i \cdot \mathbf{C}_j}{\|\mathbf{D}_i\| \|\mathbf{C}_j\|} \right) \quad \text{when} \quad i = j
\label{eq:validation}
\end{equation*}
Where:
\begin{itemize}
    \item $\mathbf{D}_i$: Is the embedding vector of the LLM-generated description for cluster $i$.
    \item $\mathbf{C}_j$: Is the centroid embedding vector of cluster $j$.
    \item $\|\mathbf{D}_i\|$: Is the magnitude of the embedding vector of $\mathbf{D}_i$.
    \item $\|\mathbf{C}_j\|$: Is the magnitude of the embedding vector of $\mathbf{C}_j$.
\end{itemize}

Accurate descriptions are expected to show the highest similarity to their respective cluster centroids, while exhibiting lower similarity to non-corresponding centroids. Cosine similarity is well-suited for the validation task 
since it focuses on how similarly two vectors point in the embedding space, rather than on their magnitude. This scale invariance ensures that even if two descriptions have different lengths or slight wording variations, as long as they convey the same underlying meaning, their vectors will still align closely \cite{Reimers2019Sentence-BERT:BERT-Networks}. Alternative measures like Euclidean distance can overemphasize differences in vector magnitude and thus misrepresent semantic proximity.



\subsection{Data Analysis}
\label{method:analysis}

\subsubsection{Thematic Disparity}

For each cluster identified, the proportion of disclosures corresponding to each agent embodiment 
was calculated. A chi-squared goodness-of-fit test was then conducted for each cluster within each experiment to determine whether these proportions significantly deviated from an even distribution across agent types, thereby evaluating whether thematic content varied depending on the embodiment of the interlocutor. This approach enables a straightforward statistical evaluation of whether thematic content distribution deviates from uniformity across agent types, providing insight into potential embodiment-driven influences on topic selection.



\subsubsection{Semantic Similarity}
To analyse the semantic similarity between disclosures to different agents of varying levels of embodiment within each cluster, we used three embedding models, including Transformers ~\cite{10.5555/3295222.3295349}, Word2Vec ~\cite{Mikolov2013DistributedCompositionality,mikolov_efficient_2013}, and BERT ~\cite{Devlin2018BERT:Understanding}.
We calculated the semantic similarity for each possible pair of agent embodiment 
within a cluster. Each embedding model provides a different 
text representation, with BERT capturing contextual meaning, Word2Vec offering a baseline for word-level similarity ~\cite{johnson2024detailed}, and Sequence Transformers focusing on syntactic structure ~\cite{selva2021review}.
One-sample t-tests were conducted to assess if the semantic similarity scores were significantly different to a defined threshold of 0.5, which we use to represent similarity above random change. 


 
\section{Results}



\subsection{Clustering Descriptions and Validation}
\label{result:clustering}

In \textbf{Experiment 1}, 
we identified 3 meaningful clusters that yielded a total within-cluster sum of squares (WCSS) of 44.97, following values of 55.15 for 1 cluster and 49.13 for 2 clusters, showing a substantial improvement in cluster compactness with each additional cluster up to three. Cluster 0 comprised 24 responses (30.77\%), Cluster 1 included 30 responses (38.46\%), and Cluster 2 included 24 responses (30.77\%). Based on the LLM interpretation of central responses, these clusters were labelled as (1) Financial Independence and Adjustment in Academic Pursuits, (2) University Life Balance and Social Integration, and (3) Insufficient Lecturer Feedback.

In \textbf{Experiment 2}, 
we identified 3 meaningful clusters that yielded a WCSS of 107.44, following values of 126.29 and 115.71 for 1 and 2 clusters, respectively, showing a substantial improvement in cluster compactness with each additional cluster up to three. Cluster 0 included 28 responses (17.28\%), Cluster 1 included 55 responses (33.95\%), and Cluster 2 included 79 responses (48.77\%). Based on the LLM interpretation of central responses, these clusters were labelled as (1) Financial Awareness and Minimal Budgeting, (2) Family Relationships and Dynamics, and (3) Social Engagement and Leisure Activities.

In \textbf{Experiment 3}, 
we identified 3 meaningful clusters that yielded a final WCSS of 234.26, following values of 275.76 and 251.77 for 1 and 2 clusters, respectively, showing a substantial improvement in cluster compactness with each additional cluster up to three. Cluster 0 comprised 162 responses (42.41\%), Cluster 1 included 87 responses (22.77\%), and Cluster 2 included 133 responses (34.82\%). Based on the LLM interpretation of central responses, these clusters were labelled as (1) Balancing Academic Demands and Social Well-being, (2) Physical Fitness and Well-being, and (3) Balancing Social Connections.

\begin{figure}[h!]
    \centering
    \includegraphics[width=1\columnwidth]{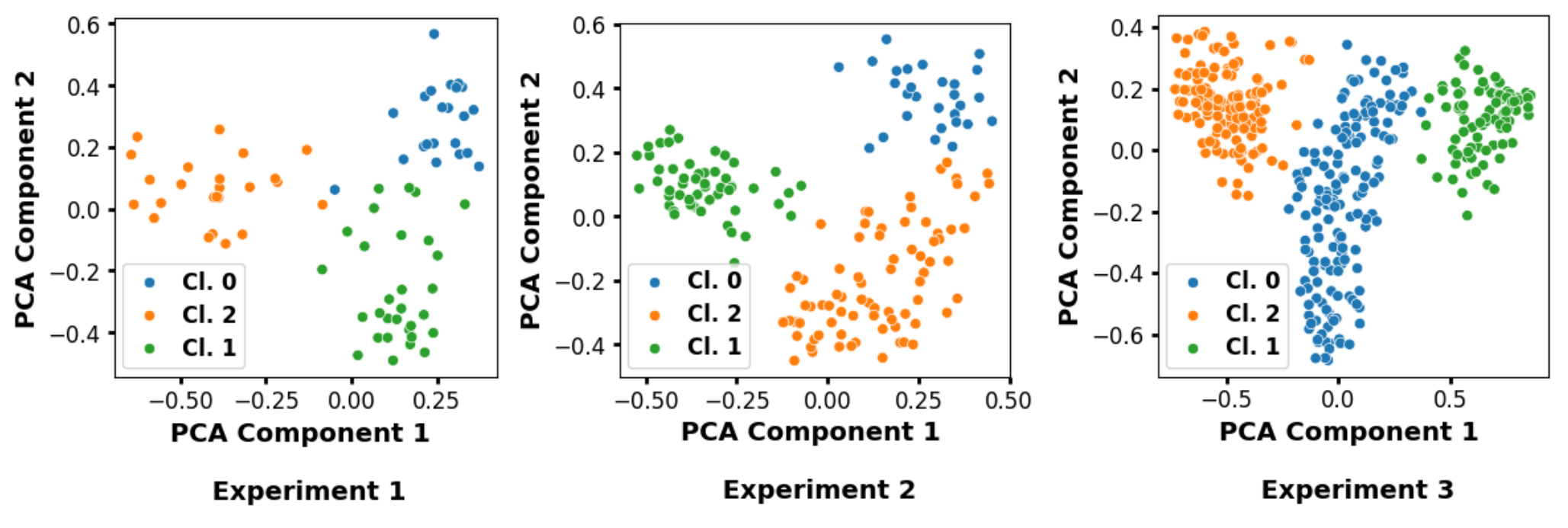}  
    \caption{Disclosures across experiments 1, 2, and 3 after clustering using PCA.}
    \label{fig:exp_clusters}
\end{figure}

Principal Component Analysis (PCA) was used to project the data onto the first two principal components, for visualizing the disclosures from all experiments, along with their corresponding cluster assignments (see Figure \ref{fig:exp_clusters}).

Across all three experiments, the assigned description for each cluster consistently showed the highest similarity to its corresponding centroid (see Section \ref{method:validation}), confirming that the LLM-generated labels captured the semantic core of each cluster (see Table \ref{table:combined_validation}). See the clusters' descriptions \href{https://osf.io/su7zq}{here}.



\begin{table}[h!]
\centering
\caption{Similarity between LLM-generated descriptions and cluster centroids across Experiments 1, 2, and 3.}
\label{table:combined_validation}
\scalebox{1}{
\begin{tabular}{|c|c|c|c|c|}
\hline
\textbf{Exp.} & \textbf{Cluster} & \textbf{Description 1} & \textbf{Description 2} & \textbf{Description 3} \\ 
\hline
\multirow{3}{*}{\textbf{1}} & 0 & \textbf{0.624} & 0.307 & 0.175 \\ 
                            & 1 & 0.473 & \textbf{0.687} & 0.352 \\ 
                            & 2 & 0.215 & 0.216 & \textbf{0.719} \\ 
\hline
\multirow{3}{*}{\textbf{2}} & 0 & \textbf{0.632} & 0.101 & 0.245 \\ 
                            & 1 & 0.074 & \textbf{0.616} & 0.351 \\ 
                            & 2 & 0.147 & 0.213 & \textbf{0.575} \\ 
\hline
\multirow{3}{*}{\textbf{3}} & 0 & \textbf{0.532} & 0.404 & 0.346 \\ 
                            & 1 & 0.253 & \textbf{0.621} & 0.212 \\ 
                            & 2 & 0.461 & 0.253 & \textbf{0.642} \\ 
\hline
\end{tabular}}
\vspace{5pt}
\end{table}

\subsection{Thematic Consistency Across Agent Embodiments}
\label{result:ratio}

In \textbf{Experiment 1}, Participants’ response types within each cluster were generally balanced. In \textit{Cluster 0}, 29\% of responses were H2H, 38\% H2R, and 33\% H2A; in \textit{Cluster 1}, 30\% were H2H, 47\% H2R, and 23\% H2A; and in \textit{Cluster 2}, 42\% were H2H, 12\% H2R, and 46\% H2A. A Chi-square goodness-of-fit test indicated that none of the clusters significantly deviated from a uniform distribution: Cluster 0, $\chi^2$(2) = 0.25, $p$ = .882; Cluster 1, $\chi^2$(2) = 2.60, $p$ = .273; Cluster 2, $\chi^2$(2) = 4.75, $p$ = .093. 

In \textbf{Experiment 2}, we found that in \textit{Cluster 0}, 43\% of responses were H2H, 21\% H2R, and 36\% H2A; in \textit{Cluster 1}, 29\% were H2H, 36\% H2R, and 35\% H2A; and in \textit{Cluster 2}, 33\% were H2H, 35\% H2R, and 32\% H2A. None of the clusters showed a significant deviation from a uniform response distribution: Cluster 0, $\chi^2$(2) = 2.00, $p$ = .368; Cluster 1, $\chi^2$(2) = 0.47, $p$ = .789; Cluster 2, $\chi^2$(2) = 0.18, $p$ = .915.

In \textbf{Experiment 3}, the proportions remained largely consistent across clusters. In \textit{Cluster 0}, 33\% of responses were H2H, 34\% H2R, and 33\% H2A; in \textit{Cluster 1}, 34\% were H2H, 39\% H2R, and 26\% H2A; and in \textit{Cluster 2}, 35\% were H2H, 31\% H2R, and 34\% H2A. Goodness-of-fit tests again revealed no significant differences from uniformity: Cluster 0, $\chi^2$(2) = 0.04, $p$ = .982; Cluster 1, $\chi^2$(2) = 2.14, $p$ = .343; Cluster 2, $\chi^2$(2) = 0.42, $p$ = .810.





\begin{table}[h!]
\centering
\caption{Raw number and proportion 
per cluster in each experiment.}
\label{table:cluster_gof}
\resizebox{\columnwidth}{!}{%
\begin{tabular}{|c|c|c|c|c|c|c|}
\hline
\textbf{Exp.} & \textbf{Cluster} & \textbf{H2H (Prop)} & \textbf{H2R (Prop)} & \textbf{H2A (Prop)} & \textbf{$\chi^2$} & \textbf{\textit{p}} \\
\hline
\multirow{3}{*}{\textbf{1}} 
& 0 & 7 (0.29) & 9 (0.38) & 8 (0.33) & 0.25 & .882 \\
& 1 & 9 (0.30) & 14 (0.47) & 7 (0.23) & 2.60 & .273 \\
& 2 & 10 (0.42) & 3 (0.12) & 11 (0.46) & 4.75 & .093 \\
\hline
\multirow{3}{*}{\textbf{2}} 
& 0 & 12 (0.43) & 6 (0.21) & 10 (0.36) & 2.00 & .368 \\
& 1 & 16 (0.29) & 20 (0.36) & 19 (0.35) & 0.47 & .789 \\
& 2 & 26 (0.33) & 28 (0.35) & 25 (0.32) & 0.18 & .915 \\
\hline
\multirow{3}{*}{\textbf{3}} 
& 0 & 53 (0.33) & 55 (0.34) & 54 (0.33) & 0.04 & .982 \\
& 1 & 30 (0.34) & 34 (0.39) & 23 (0.26) & 2.14 & .343 \\
& 2 & 47 (0.35) & 41 (0.31) & 45 (0.34) & 0.42 & .810 \\
\hline
\end{tabular}}
\vspace{5pt}
\end{table}

\subsection{Semantic Similarity Across Agent Embodiments}
\label{result:similarity}

\paragraph{\textbf{Transformer}} 
In Experiment 1, the highest similarity was found between H2R and H2A in Cluster 2 ($M = .49$, 95\%CI[.45, .53]), and the lowest in Cluster 0 for the same comparison ($M = .33$, 95\%CI[.28, .37]). In Experiment 2, similarity ranged from $M = .27$, 95\%CI[.25, .28] (H2H–H2A, Cluster 1) to $M = .37$, 95\%CI[.33, .41] (H2R–H2A, Cluster 0). Experiment 3 revealed slightly higher scores, with a peak of $M = .45$, 95\%CI[.44, .46] in Cluster 1 (H2H–H2A), and a low of $M = .32$, 95\%CI[.32, .33] in Cluster 0 (H2H–H2A). None of the comparisons reached significance above the 0.50 threshold ($p > .05$ for all), suggesting that semantic consistency, while moderate, was not statistically significant above chance for this model.

\paragraph{\textbf{Word2Vec}} The Word2Vec model yielded consistently high similarity scores across all clusters and experiments, with all comparisons significantly exceeding the 0.50 threshold ($p < .001$). In Experiment 1, scores ranged from $M = .83$, 95\%CI[.81, .85] (H2R–H2A, Cluster 0) to $M = .93$, 95\%CI[.93, .94] (H2H–H2A, Cluster 1). In Experiment 2, values ranged from $M = .83$, 95\%CI[.82, .85] (H2H–H2A, Cluster 1) to $M = .91$, 95\%CI[.90, .91] (H2H–H2R, Cluster 1). Experiment 3 further supported this trend, with all scores above 0.90, and the highest in Cluster 2 (H2H–H2A; $M = .94$, 95\%CI[.93, .94]). These results suggest robust semantic alignment in participants’ disclosures across all embodiments.

\paragraph{\textbf{BERT}} Semantic similarity scores obtained using BERT also indicated high semantic consistency across agent types. All comparisons significantly exceeded the 0.50 threshold ($p < .001$). In Experiment 1, values ranged from $M = .79$, 95\%CI[.76, .83] (H2H–H2R, Cluster 2) to $M = .85$, 95\%CI[.83, .86] (H2R–H2A, Cluster 1). In Experiment 2, scores ranged from $M = .80$, 95\%CI[.79, .80] (H2H–H2A, Cluster 1) to $M = .85$, 95\%CI[.84, .85] (H2R–H2A, Cluster 2). Experiment 3 showed similarly high consistency, with values between $M = .84$ and $M = .86$ across all comparisons. These results suggest that semantic consistency 
is robust even when accounting for contextual embeddings.

\begin{table}[h!]
\centering
\renewcommand{\arraystretch}{1.4}
\caption{Semantic similarity scores across agent embodiment pairs}
\label{table:combined_similarity}
\resizebox{\columnwidth}{!}{%
\begin{tabular}{|l|c|c|c|c|c|}
\hline
\textbf{Model}     & \textbf{Exp.} & \textbf{Comparison} & \textbf{Cluster 0} & \textbf{Cluster 1} & \textbf{Cluster 2} \\ 
\hline
\multirow{9}{*}{\textbf{Transformers}} & \multirow{3}{*}{1} & H2H vs H2A & .38 [.35, .41] & .40 [.37, .43] & .37 [.34, .39] \\ 
                                       &                    & H2H vs H2R & .34 [.30, .38] & .41 [.39, .43] & .39 [.34, .44] \\ 
                                       &                    & H2R vs H2A & .33 [.28, .37] & .42 [.39, .44] & .49 [.45, .53] \\ \cline{2-6}
                                       & \multirow{3}{*}{2} & H2H vs H2A & .36 [.34, .38] & .27 [.25, .28] & .33 [.32, .34] \\ 
                                       &                    & H2H vs H2R & .34 [.31, .37] & .31 [.29, .32] & .35 [.34, .35] \\ 
                                       &                    & H2R vs H2A & .37 [.33, .41] & .29 [.27, .30] & .32 [.31, .33] \\ \cline{2-6}
                                       & \multirow{3}{*}{3} & H2H vs H2R & .33 [.33, .34] & .41 [.40, .42] & .40 [.40, .41] \\ 
                                       &                    & H2H vs H2A & .32 [.32, .33] & .45 [.44, .46] & .41 [.40, .41] \\ 
                                       &                    & H2R vs H2A & .33 [.32, .33] & .42 [.41, .43] & .39 [.39, .40] \\ 
\hline
\multirow{9}{*}{\textbf{Word2Vec}}     & \multirow{3}{*}{1} & H2H vs H2A & .89*** [.87, .90] & .93*** [.93, .94] & .85*** [.83, .86] \\ 
                                       &                    & H2H vs H2R & .85*** [.84, .87] & .92*** [.92, .93] & .85*** [.81, .89] \\ 
                                       &                    & H2R vs H2A & .83*** [.81, .85] & .91*** [.90, .92] & .90*** [.88, .91] \\ \cline{2-6}
                                       & \multirow{3}{*}{2} & H2H vs H2A & .88*** [.86, .88] & .83*** [.82, .85] & .90*** [.90, .91] \\ 
                                       &                    & H2H vs H2R & .83*** [.81, .84] & .91*** [.90, .91] & .90*** [.90, .91] \\ 
                                       &                    & H2R vs H2A & .84*** [.83, .86] & .83*** [.82, .85] & .88*** [.88, .89] \\ \cline{2-6}
                                       & \multirow{3}{*}{3} & H2H vs H2R & .92*** [.92, .93] & .92*** [.92, .93] & .93*** [.93, .94] \\ 
                                       &                    & H2H vs H2A & .92*** [.91, .92] & .93*** [.92, .93] & .94*** [.93, .94] \\ 
                                       &                    & H2R vs H2A & .91*** [.90, .91] & .92*** [.91, .92] & .92*** [.92, .93] \\ 
\hline
\multirow{9}{*}{\textbf{BERT}}         & \multirow{3}{*}{1} & H2H vs H2A & .85*** [.84, .86] & .85*** [.83, .86] & .80*** [.79, .82] \\ 
                                       &                    & H2H vs H2R & .81*** [.79, .83] & .83*** [.82, .84] & .79*** [.76, .83] \\ 
                                       &                    & H2R vs H2A & .81*** [.79, .82] & .85*** [.84, .86] & .85*** [.83, .86] \\ \cline{2-6}
                                       & \multirow{3}{*}{2} & H2H vs H2A & .83*** [.81, .84] & .80*** [.79, .80] & .83*** [.83, .84] \\ 
                                       &                    & H2H vs H2R & .82*** [.81, .84] & .83*** [.83, .84] & .84*** [.84, .85] \\ 
                                       &                    & H2R vs H2A & .83*** [.81, .84] & .83*** [.82, .84] & .85*** [.84, .85] \\ \cline{2-6}
                                       & \multirow{3}{*}{3} & H2H vs H2R & .85*** [.84, .85] & .84*** [.84, .85] & .85*** [.84, .85] \\ 
                                       &                    & H2H vs H2A & .84*** [.84, .85] & .85*** [.84, .85] & .86*** [.86, .87] \\ 
                                       &                    & H2R vs H2A & .84*** [.84, .85] & .86*** [.86, .87] & .85*** [.85, .86] \\ 
\hline
\multicolumn{6}{l}{\textit{Note:} $p < 0.001 = ***$} \\
\end{tabular}}
\end{table}

\section{Discussion}


\subsection{Thematic Consistency Across Embodiments}
Our results revealed no significant deviations in the distribution of thematic clusters across agent types, suggesting that the content of what individuals chose to disclose was largely unaffected by the embodiment of the conversational partner. This finding supports the notion that the choice of topics in self-disclosure may be driven more by individual preferences, the structure of the interaction, or the questions asked rather than by the nature of the agent itself. It aligns with prior work suggesting that while people adapt their speech acoustically or behaviourally in HRI contexts \cite{Laban2021TellSpeech,Cohn2022Acoustic-phoneticSpeech}, communicative themes and goals may remain governed by social norms and intrinsic motivations \cite{RefWorks:455, Berg1987ThemesSelf-Disclosure}. Thematic stability across agent embodiments might also suggests that thematic decisions in self-disclosure to social robots may be guided by similar fundamental psychological mechanisms as human-human interaction. 

\subsection{Semantic Similarity Across Embodiments}

In line with the thematic consistency observed across agents, our findings reveal strong semantic similarity in the way participants expressed themselves when discussing similar topics with different interlocutors. Both Word2Vec and BERT models consistently produced high similarity scores across all agent pairs within each cluster, well above chance, suggesting that participants articulated their disclosures in comparable ways, irrespective of the agent's embodiment. 
This pattern implies that participants did not substantially adapt their lexical or semantic framing based on the agent's embodiment, instead maintaining consistent wording, phrasing, and core meanings across the three interaction conditions. While the Transformer model yielded moderately lower similarity scores, these differences likely reflect variation in how the model handles surface structure and syntactic composition, rather than substantial changes in participants’ communicative intent ~\cite{selva2021review}. 
While previous research has documented shifts in acoustic features \cite{Laban2021TellSpeech}, turn-taking \cite{10.5555/3721488.3721593}, 
or lexical changes in response to the number of participants in the interaction \cite{Asano2022ComparisonInteractions}, our results suggest that when it comes to sociolinguistic performance in self-disclosures, individuals may prioritize message consistency over stylistic adaptation.

\subsection{Implications and Considerations}

Several important considerations remain. First, the specific questions within each experiment 
may have constrained or guided participant disclosures. This design allowed for the elicitation of disclosure \cite{Laban2024StudyingParadigms}, as well as direct comparisons, but may also have limited the topics of disclosure. 
Second, the single-session nature of the interactions mean that the findings do not fully address whether deeper or more intimate disclosures might vary with repeated, longer-term engagement. Longitudinal research indicates that increasing familiarity with a robot can lead to expanded self-disclosure \cite{Laban2024BuildingTime,laban_ced_2023}; future studies should explore whether thematic variety and semantic content remains consistent under those extended circumstances. Finally, all experiments were lab-based, with carefully controlled conditions. While this yields robust cross-comparisons, 
it remains unclear whether 
contextual cues 
might influence disclosure themes more strongly.

\section{Conclusions}

 Our results provide evidence that participants’ thematic focus and semantic content remain stable across interlocutors, regardless of the agents’ embodiment. This finding underscores 
 broader social-cognitive processes in shaping the content of disclosures, rather than the physical form 
 of the conversational partner. This finding advances our theoretical understanding of interpersonal communication with social robots, but also offers practical guidance for the design of future conversational interactions with social robots and other artificial agents. Researchers and practitioners 
 should focus on creating interactions that adhere to user 
 preferences, without assuming that the agent’s embodiment alone can compel individuals to alter their social communication. 


\bibliographystyle{IEEEtran}
\balance{\bibliography{IEEEabrv,references}}

\end{document}